\documentclass{PoS}

\title{Polyakov loop renormalization with gradient flow}

\ShortTitle{Polyakov loop renormalization with gradient flow}

\author{Peter Petreczky\\
        Physics Department, Brookhaven National Laboratory, Upton, NY 11973, USA\\
        E-mail: \email{petreczk@quark.phy.bnl.gov}}

\author{\speaker{Hans-Peter Schadler}\\
				Physics Department, Brookhaven National Laboratory, Upton, NY 11973, USA\\
        Institute of Physics, University of Graz, 8010 Graz, Austria\\
        E-mail: \email{hps@abyle.org}}

\abstract{We propose to use the gradient flow for the renormalization of Polyakov loops in various representations. We study Polyakov loops in $2+1$ flavor QCD using the HISQ action and lattices with temporal extents $N_\tau$=$6$, $8$, $10$ and $12$ in various representations, including fundamental, sextet, adjoint, decuplet, 15-plet and 27-plet. This alternative renormalization procedure allows for the renormalization over a large temperature range from $T=100$~MeV -- $800$~MeV, with small errors not only for the fundamental, but also for the higher representations of the Polyakov loop. We discuss the results of this procedure and Casimir scaling of the Polyakov loop.}

\FullConference{The 33rd International Symposium on Lattice Field Theory\\
		14 -18 July 2015\\
		Kobe International Conference Center, Kobe, Japan*}

\usepackage{amsmath} 

\usepackage[utf8]{inputenc}

\newcommand{\tr}{\text{Tr}}

\begin{document}

\section{Introduction}
The theory of strong interactions, quantum chromodynamics (QCD), is one of the main building blocks of the standard model of particle physics and lattice QCD is one of the most important tools to study observables from first principle. Even though the approach has been used with great success over the last decades, there are still technical aspects which can be improved. In this work we want to discuss a new renormalization procedure for the Polyakov loop based on the gradient flow~\cite{Luscher:2010iy}. The conventional approaches depend on the calculation of additional quantities, like the static potential~\cite{Kaczmarek:2002mc}, and are therefore numerically expensive or introduce additional uncertainties which will be reflected in larger statistical and systematic errors. The gradient flow provides a direct method to renormalize the Polyakov loop without the calculation of additional quantities. We explain this approach in detail and show a comparison with results obtained from the conventional method. 
We also study the renormalized Polyakov loops in higher representations and test the so-called Casimir scaling hypothesis.

The local unrenormalized (bare) Polyakov loop on the lattice in the fundamental ``3'' representation for a spatial point ${\bf x}$ is given by
\begin{equation}\label{eq:pl}
	L^{\rm{bare}}_3({\bf x}) = \frac{1}{3} \tr \prod_{\tau=1}^{N_\tau} U_4({\bf x},\tau) \; .
\end{equation}
The matrices $U_4({\bf x},\tau)$ are elements of the group $SU(3)$ and $N_\tau$ is the temporal extent of the lattice. In actual calculations one uses the translational invariance on the lattice and averages over all spatial points
\begin{equation}\label{eq:plspatial}
	P^{\rm{bare}}_3 = \frac{1}{V_3} \sum_{{\bf x}} L^{\rm{bare}}_3({\bf x}) \; ,
\end{equation}
with the spatial lattice volume $V_3=N_s^3$. One usually considers the expectation value $\langle |P^{\rm{bare}}_3| \rangle$; from now on we will use a streamlined notation and omit the brackets and always consider this expectation value in the figures below.

The free energy of a static quark is given by the logarithm of the Polyakov loop
\begin{equation}
	F_3=-T \ln P_3 \; ,
\end{equation}
and in pure gauge theory it is infinite in the confined phase and finite in the deconfined phase above a critical temperature $T_c$. It has been studied extensively in pure gluonic theory
(see e.g. Refs.~\cite{Kaczmarek:2002mc,Kaczmarek:2004gv,Digal:2003jc,Gupta:2007ax,Mykkanen:2012ri}).
In full QCD the phase transition turns into a crossover and quark-antiquark pairs can be generated dynamically, given high enough energies. The free energy for full QCD is always finite due to color screening even below the pseudo-critical transition temperature $T_c$. This change of behavior reflects also the fact that the Polyakov loop is no longer an order parameter for the confinement/deconfinement transition and the center symmetry is not only spontaneously but also explicitly broken. However, the Polyakov loop can still serve as an observable to study the temperature dependence of screening properties of the theory (see e.g. discussions in Refs.~\cite{Bazavov:2009zn,Cheng:2007jq,Bazavov:2013yv,Borsanyi:2010bp,Borsanyi:2015yka}). 
For example for high temperatures it was found to be related to the Debye screening mass~\cite{Petreczky:2005bd}, or it can be used to study the binding properties of quarkonia at very low temperatures~\cite{Digal:2001iu}.

For such studies proper renormalization of the Polyakov loop is necessary. A unrenormalized Polyakov loop does not correspond to a physical quantity in the continuum limit and on the lattice the unrenormalized, continuum extrapolated Polyakov loop is always zero, even in the deconfinement region. Only after proper normalization a continuum limit can be defined. The renormalized Polyakov loop~\cite{Polyakov:1980ca} is given by 
\begin{equation}
	P_3(T) \equiv P^{\rm{ren}}_3(T) = e^{-c(a)N_{\tau}} P^{\rm{bare}}_3(T) \; ,
\end{equation}
where $c(a)$ is a renormalization constant which has to be determined for every lattice spacing separately  by calculating the zero temperature quark anti-quark potential. 
In addition the renormalization factor scales with the temporal lattice extent $N_{\tau}$.

In this contribution we report on the study of renormalized Polyakov loops in different representations in 2+1 flavor QCD using Symanzik flow \cite{Fodor:2014cpa} and gauge configurations generated by HotQCD collaboration 
with highly improved staggered quark (HISQ) action \cite{Bazavov:2011nk,Bazavov:2014pvz}.
A detailed description of this calculation is given in Ref. \cite{Petreczky:2015yta}.

\section{Gradient flow}

Instead of determining the renormalization constant $c(a)$, we will use the properties of the gradient flow to obtain a renormalized Polyakov loop. The defining equation for the gradient flow is
\begin{equation}\label{eq:grad}
	\dot V_t(x,\mu) = -g_0^2(\partial_{x,\mu}S[V_t])V_t(x,\mu) \; ,
\end{equation}
where $g_0$ is the bare gauge coupling and $t$ (dimension $[a^2]$) is a new parameter for the evolution in flow time. The initial condition for the fields $V_t(x,\mu)$ at a lattice point $x=({\bf x}, \tau)$ in direction $\mu$ is given by
\begin{equation}
	V_t(x,\mu)|_{t=0} = U_{\mu}(x) \; .
\end{equation}
The gradient flow smears the original field $U_{\mu}(x)$ at the length scale of 
\begin{equation}
	f=\sqrt{8t} \; ,
\end{equation}
and removes the UV singularities. Therefore, operators which are evaluated at non-zero flow time do not require additional renormalization~\cite{Luscher:2011bx}. For the renormalization of the Polyakov loop this means that we can obtain the renormalized quantity by replacing the links $U_\mu(x)$ in Eq.~(\ref{eq:pl}) with the fields evolved in the flow time $V_t(x,\mu)|_{t>0}$. The choice of flow time $f=a\sqrt{8t}$ (constant in physical units fm) corresponds to a particular renormalization scheme as long as we fulfill the requirement that $a \ll f \ll 1/T$.

\begin{figure}[ht]
	\centering
	\hspace{-0.05\textwidth}
	\includegraphics[width=0.49\textwidth,clip]{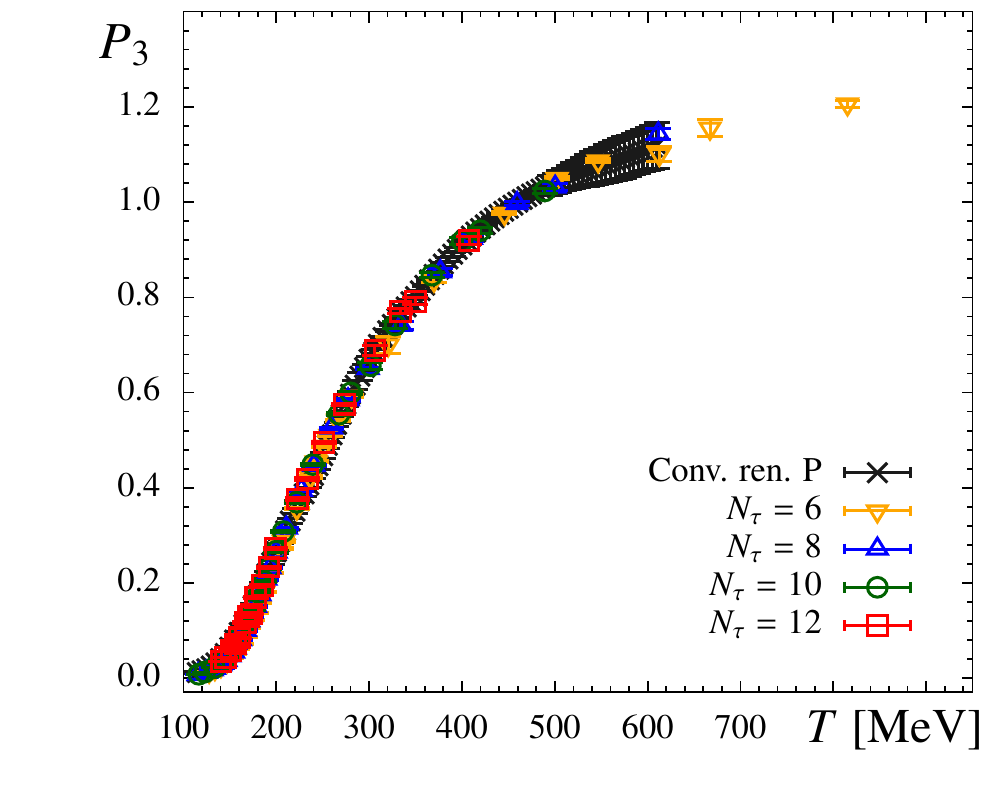}
	\hspace{0.005\textwidth}
	\includegraphics[width=0.49\textwidth,clip]{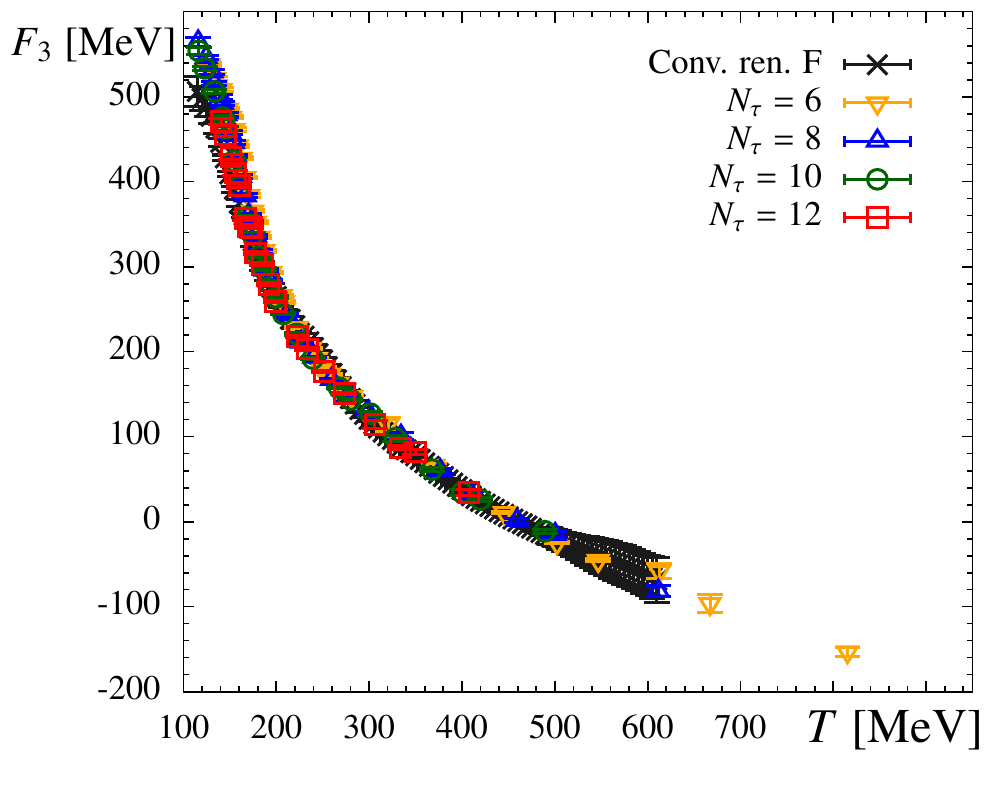}
	\caption{Renormalized fundamental Polyakov loop $P_3$ (l.h.s.) and free energy $F_3$ (r.h.s.) as a function of the temperature $T$. We show different temporal extents $N_\tau$ and compare it to the continuum extrapolated, renormalized results from~\cite{peterpolldata} (black crosses) which were obtained from the static potential. We set the renormalization scale by matching $F_3$ at $T\approx 200$~MeV with the conventionally renormalized free energy.}
	\label{fig:pollren}
\end{figure}

As we want to study a broad temperature range and we are working at finite lattice spacing, 
we have to define different flow regions. These are given by
\begin{equation}\label{eq:flowdef}
	f = \left\{ \begin{array}{rl}
	3.00 f_0 & \; \mbox{for} \;\; T \, < \, 200~{\rm MeV} \; ,\\
	2.00 f_0 & \; \mbox{for} \;\; 200~{\rm MeV} \leq T \, \leq \, 300~{\rm MeV} \; ,\\
	0.50 f_0 & \; \mbox{for} \;\; 300~{\rm MeV} \leq T \, < \, 600~{\rm MeV} \; ,\\
	0.25 f_0 & \; \mbox{for} \;\; T \, \geq \, 600~{\rm MeV} \; ,\\
	\end{array} \right.
\end{equation}
where we state the value of $f$ in units of $f_0 = 0.2129$~fm. As the different flow times are ideally related by a constant shift of the free energy, we match the different regions by determining the difference of the free energies obtained at an overlapping temperature point. These offsets are determined separately for every temporal extent $N_\tau$ and used to match the different flow regions. The temperature points at which these shifts are determined were chosen to be always the data point just below the new flow region, e.g. for $N_\tau=12$ the matching temperatures are $T=199$~MeV, $273$~MeV and $563$~MeV.

The results of this procedure are shown in Fig.~\ref{fig:pollren}. We plot the Polyakov loop $P_3$ (l.h.s.) and the free energy $F_3$ (r.h.s.) in the fundamental representation as a function of the temperature for different temporal extents $N_\tau$. The black crosses are continuum extrapolated, renormalized results from~\cite{peterpolldata} obtained by using the static potential at zero temperature. To set the renormalization scale we match $F_3$ at $T\approx 200$~MeV with the conventionally renormalized free energy for the ensemble with temporal extent $N_\tau=12$. For $N_\tau=6$, $8$ and $10$ we use the same constant shift, which guarantees that the cut-off effects from the different $N_\tau$ are not obscured. From these figures it is clear that this approach reproduces the conventional results up to a constant shift. Small cut-off effects are visible, but in general even the non-continuum extrapolated results already agree with the conventionally renormalized Polyakov loop.

Judging from this comparison, we can conclude that the gradient flow renormalization approach works and reproduces the regular renormalized Polyakov loop up to a constant shift of the free energy. This shift comes from the difference in the renormalization scheme and different approaches can always be matched by determining this shift. Now that the agreement between the methods is established, we want to use the gradient flow renormalization method to calculate higher representations of the Polyakov loop.

\section{Higher representations and Casimir scaling}

\begin{figure}[ht]
	\centering
	\hspace{-0.05\textwidth}
	\includegraphics[width=0.46\textwidth,clip]{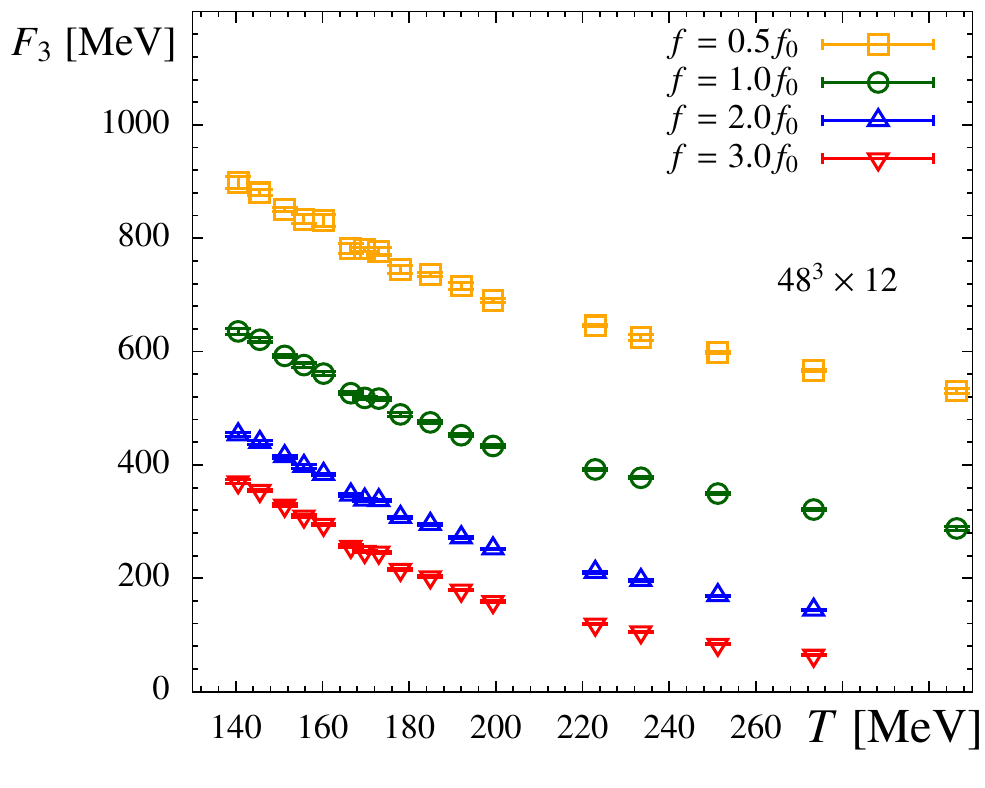}
	\hspace{0.005\textwidth}
	\includegraphics[width=0.46\textwidth,clip]{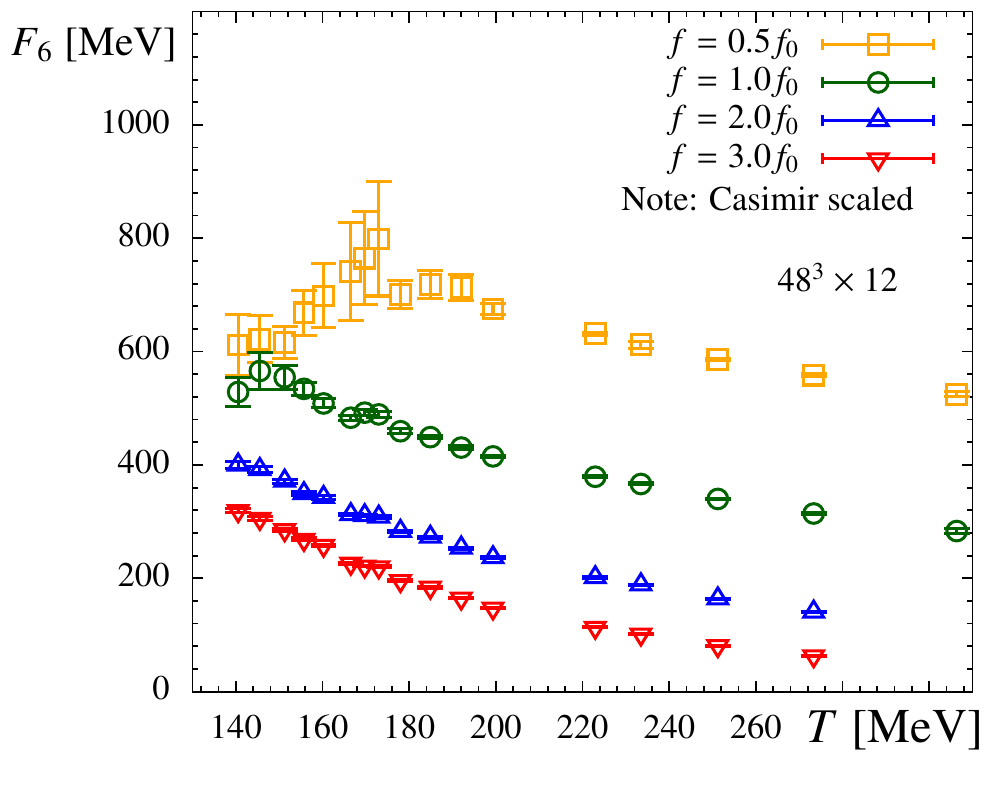}\\
	\hspace{-0.05\textwidth}
	\includegraphics[width=0.46\textwidth,clip]{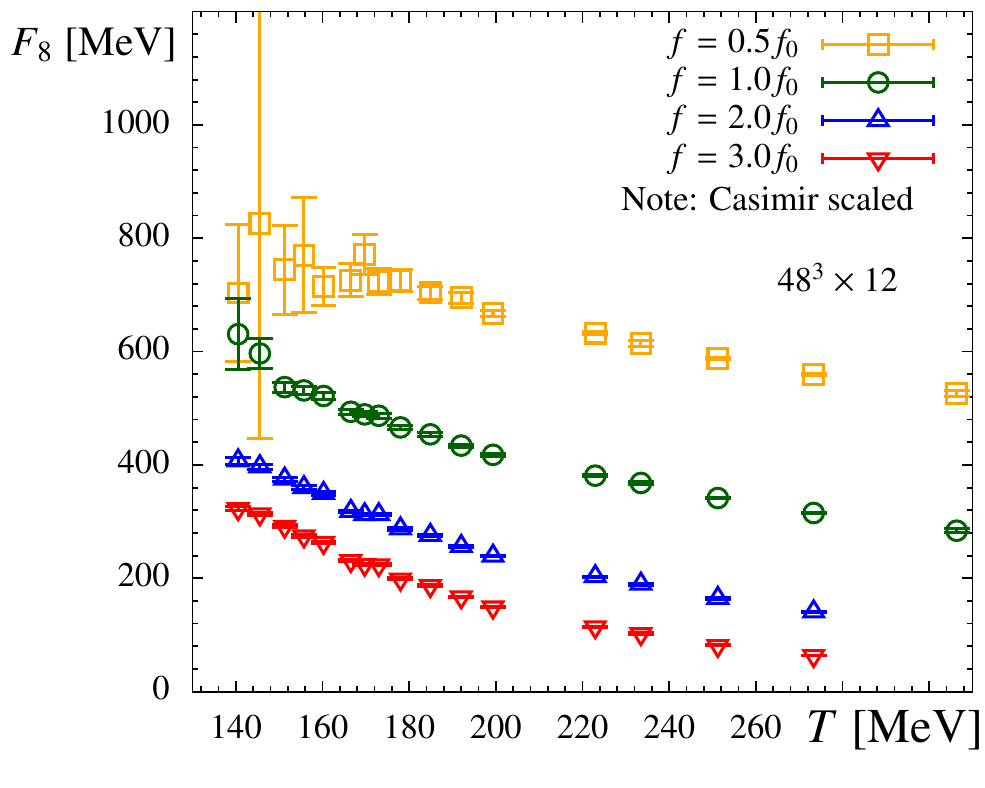}
	\hspace{0.005\textwidth}
	\includegraphics[width=0.46\textwidth,clip]{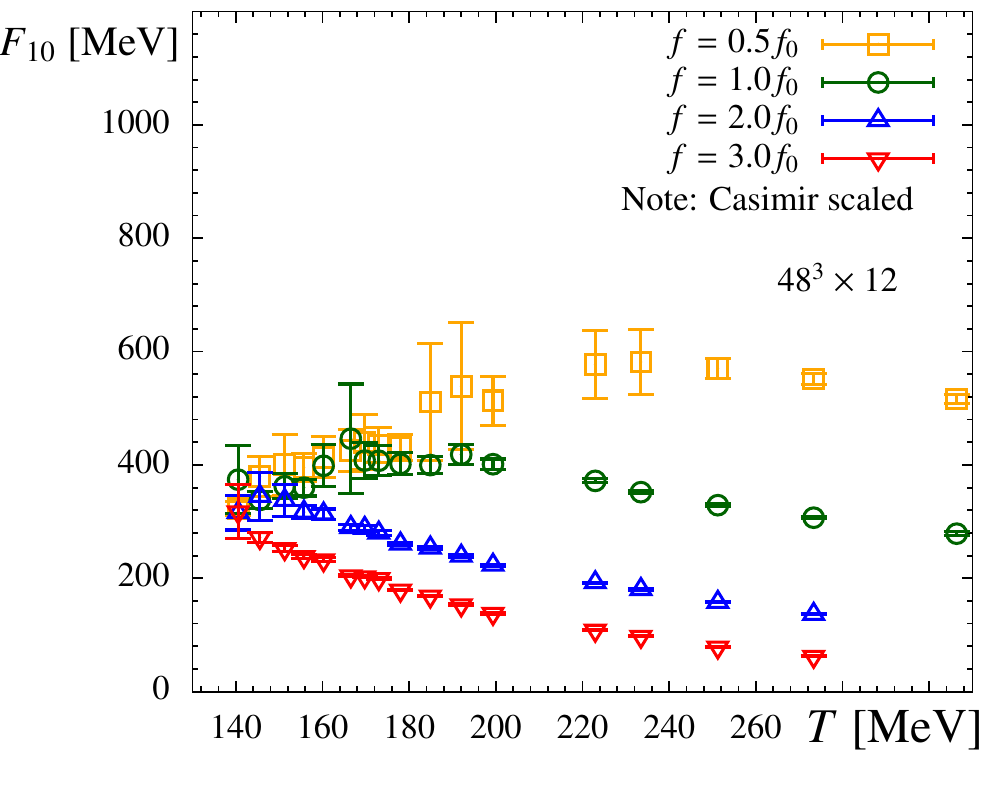}
	\caption{Free energy in representations $3$, $6$, $8$ and $10$ as a function of the temperature $T$ for different flow times $f$.}
	\label{fig:higherrep1}
\end{figure}

\begin{figure}[ht]
	\centering
	\hspace{-0.05\textwidth}
	\includegraphics[width=0.46\textwidth,clip]{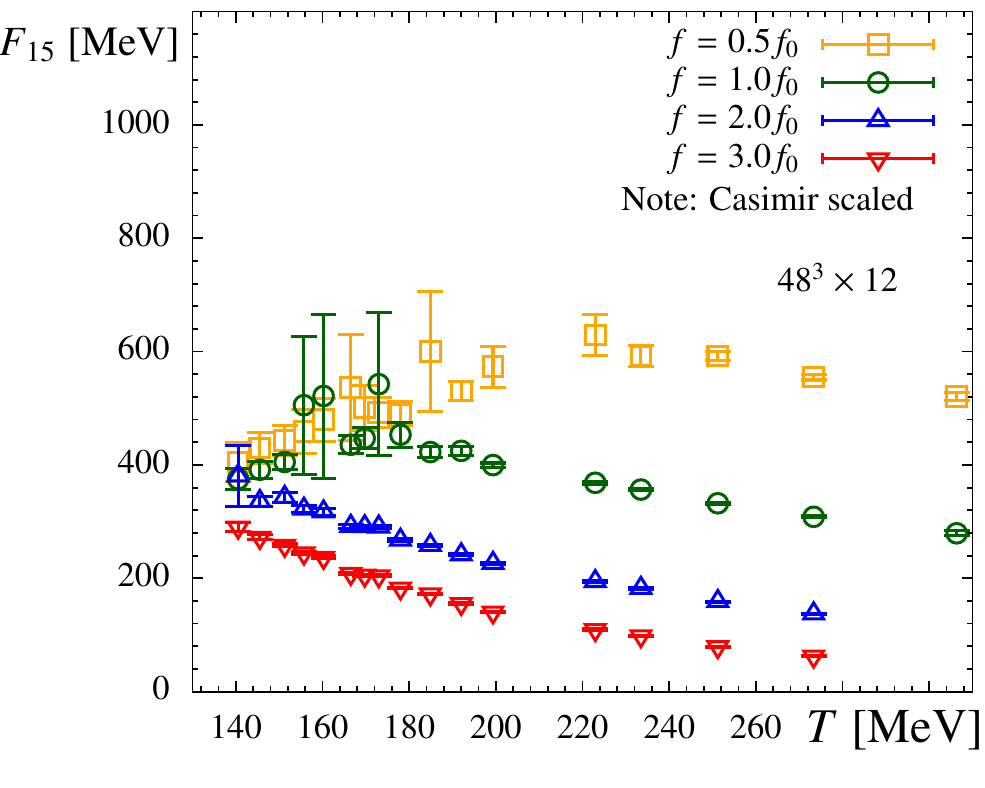}
	\hspace{0.005\textwidth}
	\includegraphics[width=0.46\textwidth,clip]{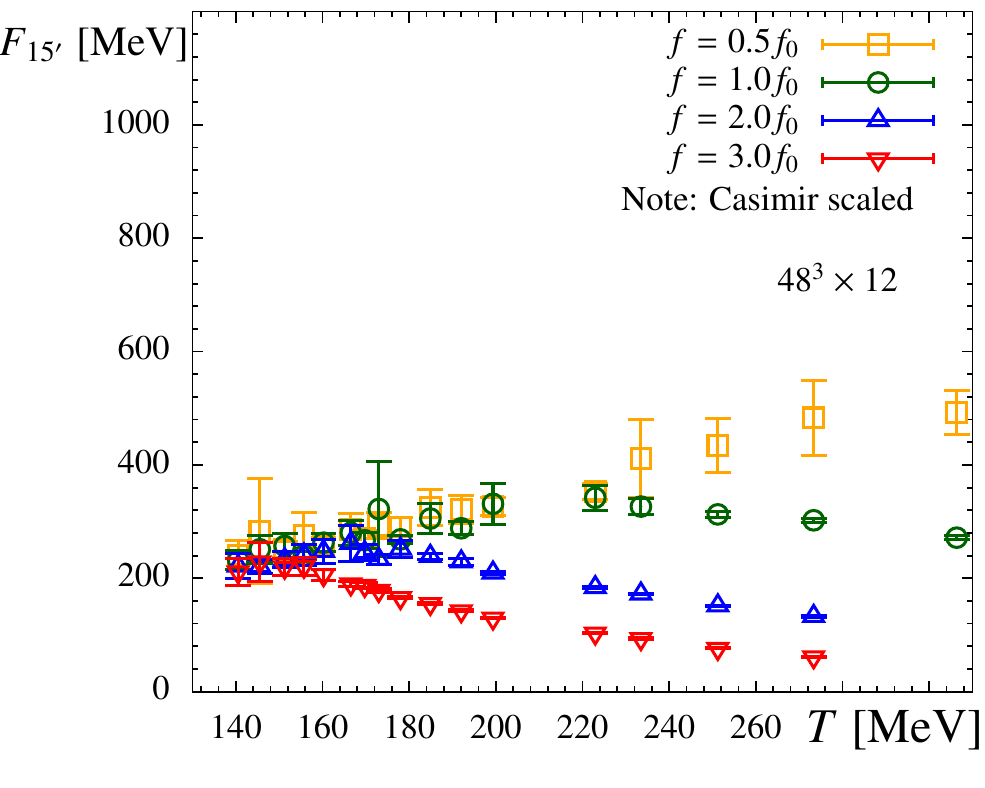}\\
  \hspace{-0.05\textwidth}
	\includegraphics[width=0.46\textwidth,clip]{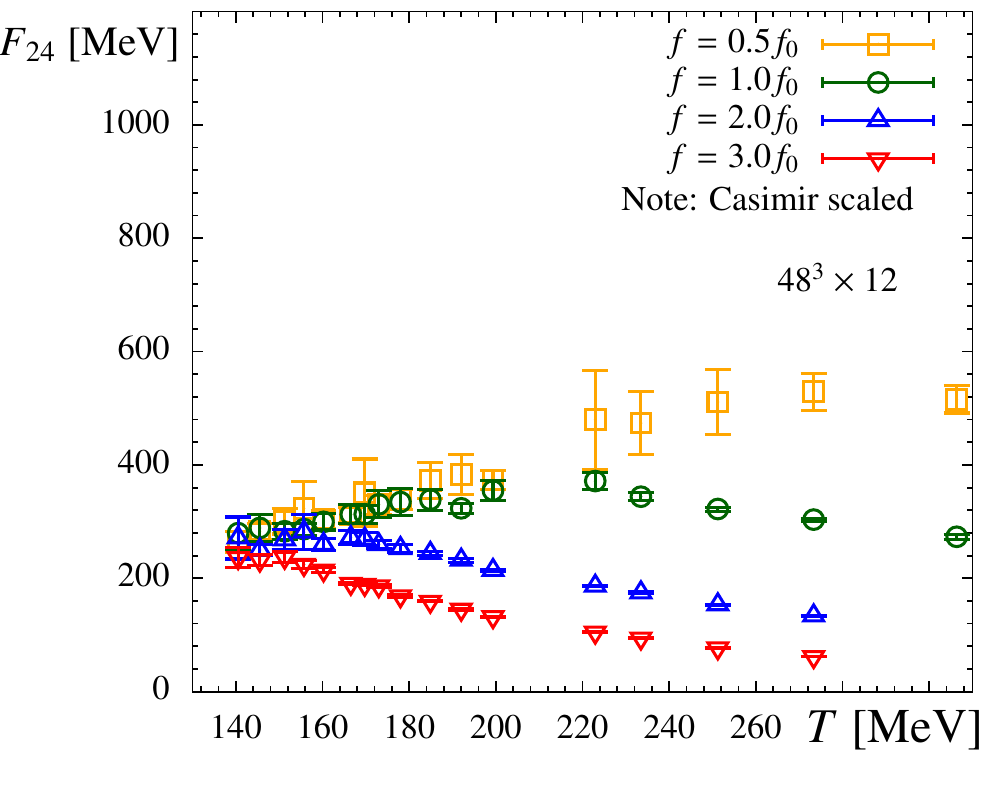}
	\hspace{0.005\textwidth}
	\includegraphics[width=0.46\textwidth,clip]{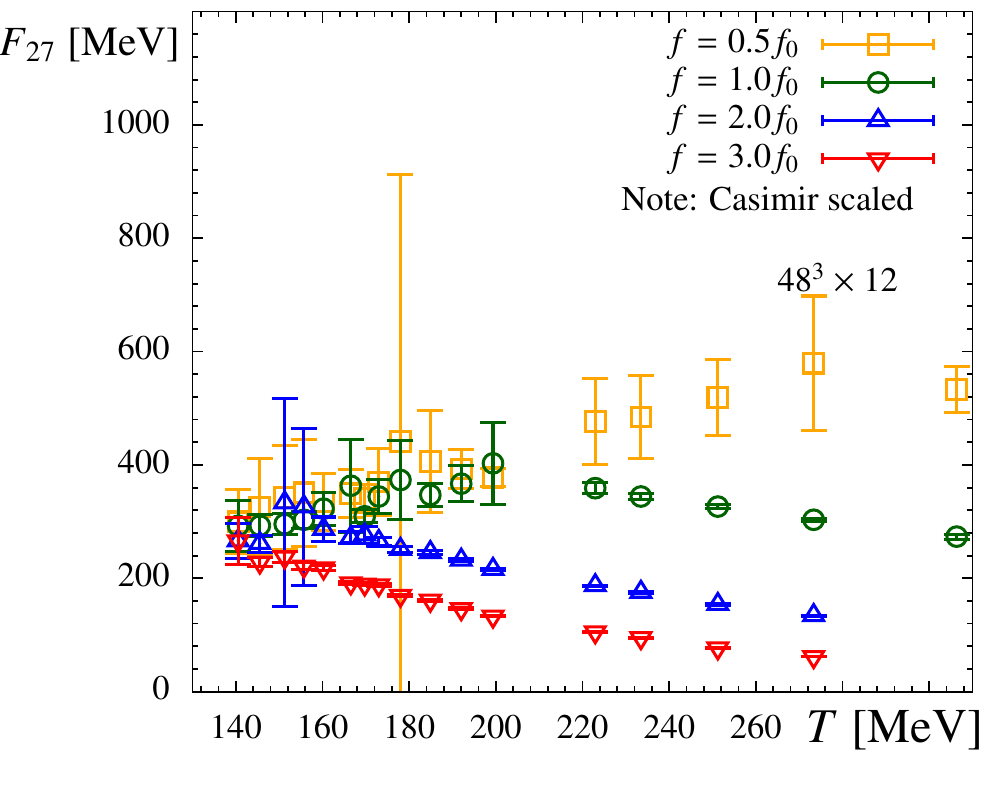}
	\caption{Free energy in representations $15$, $15'$, $24$ and $27$ as a function of the temperature $T$ for different flow times $f$.}
	\label{fig:higherrep2}
\end{figure}

In addition to the Polyakov loop in the fundamental representation, we also consider the higher representations $N=6$, $8$, $10$, $15$, $15'$, $24$ and $27$. These can be constructed from the local Polyakov loops in the fundamental representation using group theory relations as follows (see Ref.~\cite{Gupta:2007ax} for details):
\begin{align}
	L_6&=\frac{1}{6}(l_3^2-l_3^*) \; , \; &L_{15'}&=\frac{1}{15}(l_3 l_{10}-l_{15})  \; ,\notag\\
	L_8&=\frac{1}{8}(|l_3|^2-1) \; , \; &L_{24}&=\frac{1}{24}(l_3^* l_{10}-l_{6})  \; ,\notag\\
	L_{10}&=\frac{1}{10}( l_3 l_6-l_8) \; , \; &L_{27}&=\frac{1}{27}(|l_6|^2-l_{8}-1)  \; ,\notag\\
	L_{15}&=\frac{1}{15}(l_3^* l_6-l_3) \; ,
\end{align}
where $l_3=3 L_3$ and $l_3^*$ is its complex conjugate.

Casimir scaling means that the free energy of a static charge in representation $N$ is proportional
to the quadratic Casimir operator $C_N$ of that representation.
If Casimir scaling holds, for the Polyakov loop  we have
\begin{equation}
	(P_3)^{1/d_3} = (P_6)^{1/d_6} = (P_8)^{1/d_8} = ... \; ,
\end{equation}
with $d_N=C_2(N)/C_2(3)$ the ratio of the quadratic Casimir operators 
of the fundamental and the $N$-representation. Stated differently, if Casimir scaling holds $(P_N)^{1/d_N}$ is independent of the representation $N$.

We use this relation to plot the free energy in different representations as a function of the temperature in Figs.~\ref{fig:higherrep1}~and~\ref{fig:higherrep2} for different flow times $f$. 
These plots show, that the gradient flow helps to extract the higher representations even for low temperatures. Especially the highest representations show artifacts for small (and zero) 
flow time which can be lifted by increasing the flow. 
In the ideal case different flow times should agree up to a constant shift. This can be observed in the figures for
sufficiently large value of $f$. We need values of $f$ larger than $2f_0$ for $T<220$~MeV and larger than $f_0$ for
$T>220$~MeV to obtain reliable results for the free energies in higher representations. For these values of the flow
time and $T>220$~MeV the Casimir scaled free energy in higher representation has the same temperature dependence as $F_3$, i.e. Casimir scaling holds.

\section{Summary}
Using gauge configurations in 2+1 flavor QCD, generated by the HotQCD collaboration with HISQ action, we studied the Polyakov 
loops in various representations.
We showed that the gradient flow provides us with an excellent tool to obtain the renormalized Polyakov loop on the lattice. We compared our results with the Polyakov loop obtained with a different method and found very good agreement. In addition we could also extract the renormalized Polyakov loop in higher representations at low temperature which is challenging in the conventional approach.

\vskip3mm
\noindent
{\bf Acknowledgements: }  
H.-P.~Schadler is funded by the FWF DK W1203 ``{\sl Hadrons in Vacuum, Nuclei and Stars}''.   
The authors want to thank Johannes Weber for interesting discussions. The numerical computations have been carried out on clusters of the USQCD collaboration, on the Vienna Scientific Cluster (VSC) and at NERSC with the publicly available MILC code~\cite{MILC}.


\begin{thebibliography}{99}

\bibitem{Luscher:2010iy} 
  M.~L\"uscher, 
  JHEP {\bf 1008}, 071 (2010), 
  JHEP {\bf 1403}, 092 (2014), 
  \href{http://arxiv.org/abs/1006.4518}{[arXiv:1006.4518 [hep-lat]]}.
  
\bibitem{Kaczmarek:2002mc} 
  O.~Kaczmarek, F.~Karsch, P.~Petreczky, F.~Zantow, 
  Phys.\ Lett.\ B {\bf 543}, 41 (2002), 
  \href{http://arxiv.org/abs/hep-lat/0207002}{[arXiv:hep-lat/0207002]}.  

\bibitem{Kaczmarek:2004gv} 
  O.~Kaczmarek, F.~Karsch, F.~Zantow and P.~Petreczky,
  Phys.\ Rev.\ D {\bf 70}, 074505 (2004), 
  [Phys.\ Rev.\ D {\bf 72}, 059903 (2005)], 
  \href{http://arxiv.org/abs/hep-lat/0406036}{[arXiv:hep-lat/0406036]}.

\bibitem{Digal:2003jc} 
  S.~Digal, S.~Fortunato and P.~Petreczky,
  Phys.\ Rev.\ D {\bf 68}, 034008 (2003), 
  \href{http://arxiv.org/abs/hep-lat/0304017}{[arXiv:hep-lat/0304017]}.

\bibitem{Gupta:2007ax} 
  S.~Gupta, K.~Huebner, O.~Kaczmarek,
  Phys.\ Rev.\ D {\bf 77}, 034503 (2008), 
  \href{http://arxiv.org/abs/0711.2251}{[arXiv:0711.2251 [hep-lat]]}.

\bibitem{Mykkanen:2012ri} 
  A.~Mykkanen, M.~Panero, K.~Rummukainen, 
  JHEP {\bf 1205}, 069 (2012), 
  \href{http://arxiv.org/abs/1202.2762}{[arXiv:1202.2762 [hep-lat]]}.

\bibitem{Bazavov:2009zn} 
  A.~Bazavov {\it et al.},
  Phys.\ Rev.\ D {\bf 80}, 014504 (2009), 
  \href{http://arxiv.org/abs/0903.4379}{[arXiv:0903.4379 [hep-lat]]}.

\bibitem{Cheng:2007jq} 
  M.~Cheng {\it et al.},
  Phys.\ Rev.\ D {\bf 77}, 014511 (2008), 
  \href{http://arxiv.org/abs/0710.0354}{[arXiv:0710.0354 [hep-lat]]}.

\bibitem{Bazavov:2013yv} 
  A.~Bazavov and P.~Petreczky,
  Phys.\ Rev.\ D {\bf 87}, 094505 (2013), 
  \href{http://arxiv.org/abs/1301.3943}{[arXiv:1301.3943 [hep-lat]]}.

\bibitem{Borsanyi:2010bp} 
  S.~Bors\'anyi {\it et al.} [Wuppertal-Budapest Collaboration],
  JHEP {\bf 1009}, 073 (2010), 
  \href{http://arxiv.org/abs/1005.3508}{[arXiv:1005.3508 [hep-lat]]}.

\bibitem{Borsanyi:2015yka} 
  S.~Bors\'anyi, Z.~Fodor, S.~D.~Katz, A.~P\'asztor, K.~K.~Szab\'o and C.~T\"or\"ok,
  JHEP {\bf 1504}, 138 (2015), 
  \href{http://arxiv.org/abs/1501.02173}{[arXiv:1501.02173 [hep-lat]]}.
  
\bibitem{Petreczky:2005bd} 
  P.~Petreczky, 
  Eur.\ Phys.\ J.\ C {\bf 43}, 51 (2005), 
  \href{http://arxiv.org/abs/hep-lat/0502008}{[arXiv:hep-lat/0502008]}.

\bibitem{Digal:2001iu} 
  S.~Digal, P.~Petreczky, H.~Satz, 
  Phys.\ Lett.\ B {\bf 514}, 57 (2001), 
  \href{http://arxiv.org/abs/hep-ph/0105234}{[arXiv:hep-ph/0105234]}.  

\bibitem{Polyakov:1980ca} 
  A.~M.~Polyakov,
  Nucl.\ Phys.\ B {\bf 164}, 171 (1980).
  
\bibitem{Fodor:2014cpa} 
  Z.~Fodor, K.~Holland, J.~Kuti, S.~Mondal, D.~Nogradi, C.~H.~Wong, 
  JHEP {\bf 1409}, 018 (2014), 
  \href{http://arxiv.org/abs/1406.0827}{[arXiv:1406.0827 [hep-lat]]}.

\bibitem{Bazavov:2011nk} 
  A.~Bazavov {\it et al.},
  Phys.\ Rev.\ D {\bf 85}, 054503 (2012), 
  \href{http://arxiv.org/abs/1111.1710}{[arXiv:1111.1710 [hep-lat]]}.

\bibitem{Bazavov:2014pvz} 
  A.~Bazavov {\it et al.} [HotQCD Collaboration],
  Phys.\ Rev.\ D {\bf 90}, 094503 (2014), 
  \href{http://arxiv.org/abs/1407.6387}{[arXiv:1407.6387 [hep-lat]]}.

\bibitem{Petreczky:2015yta} 
  P.~Petreczky and H.-P.~Schadler,
  \href{http://arxiv.org/abs/1509.07874}{arXiv:1509.07874 [hep-lat]}.
  
\bibitem{Luscher:2011bx} 
  M.~L\"uscher and P.~Weisz,
  JHEP {\bf 1102}, 051 (2011), 
  \href{http://arxiv.org/abs/1101.0963}{[arXiv:1101.0963 [hep-th]]}.

\bibitem{peterpolldata}
	Conventionally renormalized Polyakov loop data provided by P.~Petreczky and J.~Weber.

\bibitem{MILC}
  MILC collaboration, \href{http://physics.utah.edu/~detar/milc/index.html}{http://physics.utah.edu/\textasciitilde detar/milc/index.html}  



\end{thebibliography}
\end{document}